\documentstyle[12pt]{article}
\def\bkR{{\rm I\kern-.17em R}}
\def\bkC{{\rm \kern.24em \vrule width.05em height1.4ex depth-.05ex \kern-.26em C}}

\begin{document}

\markboth{Authors' Names}
{Instructions for Typing Manuscripts (Paper's Title)}

\author{Nuno Costa Dias\footnote{{\it ncdias@mail.telepac.pt}} \\ Jo\~{a}o Nuno Prata\footnote{{\it joao.prata@ulusofona.pt}} \\ {\it Departamento de Matem\'atica} \\
{\it Universidade Lus\'ofona de Humanidades e Tecnologias} \\ {\it Av. Campo Grande, 376, 1749-024 Lisboa, Portugal}}

\title{Comment on "On infinite walls in deformation quantization"}

\maketitle
\begin{abstract}
We discuss a recent method proposed by Kryukov and Walton to address boundary-value problems in the context of deformation quantization. We compare their method with our own approach and establish a connection between the two formalisms.     
\end{abstract}

\section{Introduction}

In a recent paper \cite{Walton} Kryukov and Walton have addressed the problem of infinite walls in the context of deformation quantization (see also \cite{Walton2}). They proposed a method alternative to ours \cite{Dias1,Dias2}. In this comment, we would like to compare the two approaches and prove their equivalence under some simplifying conditions.

Without loss of generality, we shall henceforth consider a one-dimensional system on the half-line $x <0$, whose state is described by a wavefunction satisfying the Dirichlet boundary condition
\begin{equation}
\psi (0)=0.
\end{equation}
As in \cite{Walton} we shall take units such that $\hbar=2m=1$. For simplicity, we shall restrict our analysis to the free particle.

In standard quantum mechanics, one solves the time independent Schr\"odinger equation,
\begin{equation}
- \psi'' (x) = E \psi (x),
\end{equation}
on the whole line, imposes the condition (1) and confines the wavefunction thereafter:
\begin{equation}
\phi (x) = \theta (-x) \psi (x).
\end{equation}
In our work we proved that the corresponding Wigner function
\begin{equation}
f(x,p)= \frac{1}{\pi} \int_{- \infty}^{+ \infty} dy \hspace{0.2 cm} e^{-2ipy} \phi^* (x-y) \phi (x+y)= \theta(-x) \frac{1}{\pi} \int_x^{-x} dy \hspace{0.2 cm} e^{-2ipy} \psi^* (x-y) \psi (x+y)
\end{equation}
does not obey the expected $\star$-genvalue equation:
\begin{equation}
p^2 \star f(x,p) = E f(x,p).
\end{equation}
Since this equation is the Weyl transform of the eigenvalue equation:
\begin{equation}
\hat{p}^2|\phi><\phi|=E|\phi><\phi|
\end{equation}
one is led to the conclusion that either the Weyl transform or the eigenvalue equation (6) fail to be valid in the confined case. Let us discuss the first possibility in some detail. Standard operator quantum mechanics provides two main (non-equivalent) formulations of the confined particle subject to Dirichlet boundary conditions, \cite{Piotr,Isham}. In the first one the confined system is formulated on the Hilbert space $L^2(I)$, where $I$ is the support of the wave function (typically the half-line or a compact interval). In this case the observables' algebra is no longer the Heisenberg algebra (notice that there is no self-adjoint representation of the Heisenberg algebra on $L^2(I)$ which is compatible with Dirichlet boundary conditions) \cite{Isham}. Since the Weyl transform only applies to the Heisenberg algebra, it cannot be used in this case. 
Another possibility is to formulate the system on the Hilbert space $L^2(\bkR)$ (i.e. acknowledging the rest of the real line) in which case the observable's algebra is the standard Heisenberg algebra and the confinement of the particle is a consequence of a suitable potential added to the Hamiltonian. In this case we say that the confinement is dynamical and the Weyl map is valid. The problem with this approach is that there is no regular potential able to confine completely the particle. We then have two possibilities. We either i) provide a family of regular potentials displaying the confined solution in the limit case or ii) we introduce a non-regular, distributional potential.  

In both papers \cite{Walton,Dias1} the authors took the dynamical confining point of view. The two approaches differ precisely in the way the boundary potential is modelled. 
In ref. \cite{Walton} the authors consider a family of regular exponential potentials $V_{\alpha}(x)=e^{2\alpha x}$ and were left with the energy eigenvalue equation:
\begin{equation}
\left( \hat p^2+V_{\alpha}(x) \right) \xi_{\alpha}(x)= E \xi_{\alpha}(x)
\end{equation}
whose solutions satisfy: $\lim_{\alpha \to \infty} \xi_{\alpha}(x) = \phi (x)$.

Applying the Weyl map to (7) we get:
\begin{equation}
\left\{ \begin{array}{l}
\left(p^2 + e^{2 \alpha x} \right) \star f_{\alpha} (x,p) = E f_{\alpha} (x,p) \\
f_{\alpha} (x,p) \star \left(p^2 + e^{2 \alpha x} \right) = E f_{\alpha} (x,p)
\end{array} \right.
\end{equation}
whose solutions (as pointed out in \cite{Walton,Walton2}) also display a well defined limit $\alpha \to \infty$:
\begin{equation}
\lim_{\alpha \to \infty}f_{\alpha}(x,p) = f(x,p)
\end{equation}
This means that if $W$ is the Weyl map then $\lim_{\alpha \to \infty} W(\xi_{\alpha})=W(\lim_{\alpha \to \infty}\xi_{\alpha})$ a property that  
reinforces our previous argument that the Weyl map properly extends to the confined case provided "the rest of the real line" is taken into account. 
Unfortunately, the pair of equations (8) (as well as eq.(7)) do not make sense in the limit $\alpha \to \infty$ given the divergent nature of the potential.
The authors of \cite{Walton} were able to manipulate eq.(8) and derive a new equation, alternative (but not equivalent) to (8) and displaying the following form in the limit $\alpha \to \infty$:
\begin{equation}
0 = \frac{1}{16} \partial_x^4 \rho (x,p) + \frac{1}{2} (p^2 +E) \partial_x^2 \rho (x,p) + (p^4 -2 E p^2 + E^2 ) \rho (x,p) .
\end{equation}
Moreover they proved that this equation is equivalent to:
\begin{equation}
(p^2-E) \star \rho (x,p) \star (p^2-E)=0
\end{equation}
and displays a solution which is related to the confined Wigner function (4) by $f(x,p)=\theta(-x) \rho(x,p)$.

Our purpose in \cite{Dias1} was also to derive a modified $\star$-genvalue equation satisfied by (4). However, we followed a completely different procedure:
First we proved that $\phi (x)$ (cf.(3)) obeys the following equation on the whole line:
\begin{equation}
- \phi'' (x) = E \phi (x) - \delta_-' (x) \phi (x),
\end{equation}
where the generalized distribution $\delta_- (x)$ is defined by:
\begin{equation}
\int dx \hspace{0.2 cm} \delta_- (x) \tilde t(x) = \lim_{\epsilon \to 0^+} \int dx \hspace{0.2 cm} \delta (x) \tilde t(x -  \epsilon),
\end{equation}
for suitable test distributions $\tilde t(x)$ \cite{Dias1,Dias2}.

Conversely, it was shown that, if $\phi$ is a solution of (12) and obeys the boundary condition 
\begin{equation}
\phi (0^-) =0,
\end{equation}
then $\phi$ takes the form (3), where $\psi$ is a solution of (1,2). Eq.(12) defines a modified Hamiltonian:
\begin{equation} 
H= p^2 + \delta_-' (x).
\end{equation}
which displays a distributional potential.
Applying the Weyl map we get the new $\star$-genvalue equations:
\begin{equation}
\left( p^2 + \delta_-' (x) \right) \star f(x,p) = E f(x,p), \hspace{0.5 cm} f(x,p)\star \left( p^2 + \delta_-' (x) \right) = E f(x,p).
\end{equation}
which constitute a system of integral-differential equations for the Wigner function. Notice that the integral term comes from the delta distribution: $\delta_-' (x)  \star f(x,p)=\lim_{\epsilon \to 0^+} \frac{1}{2\pi} \int dk \, ik \, e^{ikx} f(x-\epsilon,p-k/2)$, \cite{Dias1}. 
 
We also derived the boundary conditions to be satisfied by $f(x,p)$. Applying the Weyl map to eq.(14) we got:
\begin{equation}
\lim_{x \to 0^-} \int dp \hspace{0.2 cm} f(x,p) =0,
\end{equation}
Finally, we proved the converse result in the case of a non-degenerate spectrum. If $f(x,p)$ is a solution of (16), and satisfies the boundary condition (17)
then $f(x,p)$ takes the form (4) where the corresponding wavefunction is a solution of (12,14).\\

Several comments are now in order:\\

1) The Kryukov-Walton approach \cite{Walton} displays several interesting features. It formulates the eigenvalue problem in terms of a regular differential equation and it proves that through a suitable manipulation the limit $\alpha \to \infty$ can be consistently performed in the deformation context. Contrary to our own approach one is not required to deal with a differential equation with distributional coefficients coming from a distributional potential. This avoids having to deal, in the deformation context, with an integral-differential formulation of the eigenvalue problem. Furthermore the Kryukov-Walton approach is closer to the spirit of standard local boundary-value problems. Indeed, in the latter context one solves a local differential equation, imposes suitable boundary conditions and confines the domain thereafter (cf.(2,3)).
\\    

2) On the other hand in \cite{Walton} one loses contact with the standard operator formulation of the eigenvalue problem. Strictly speaking eq.(10) cannot be regarded as the Weyl transform of a confined version of the eigenvalue equation in standard operator quantum mechanics. In fact its solutions are not the Weyl transform of the confined eigenfunctions. Instead, some of its solutions are related to the Weyl transform of the confined eigenfunctions by $f(x,p)=\theta(x) \rho(x,p)$. However, equation (10) has many solutions. Kryukov and Walton did not provide the conditions that select the ones related to the confined stargenfunctions. This is related to the problem of extending Baker's converse construction for this formulation (i.e. the proof that if $\rho(x,p)$ satisfies eq.(10) and some possible extra conditions still to be determined then $\theta(x) \rho(x,p)$ is the Weyl transform of the corresponding confined energy eigenstate). It is also not clear how can eq.(10) be extended to non-diagonal stargenfunctions nor if a similar approach can be used to derive the dynamics of the confined system.\\

3) In \cite{Dias1} the confined eigenvalue problem is not formulated in terms of a regular potential nor as the limit of a family of regular potentials. Instead the boundary potential is a distribution yielding a formulation that makes sense quantum mechanically (i.e. it leads to the expected solutions) but which is difficult to see how it can be derived by quantizing a classical formulation of the confined particle. In this sense the modification of the Hamiltonian introduced in \cite{Dias1} is not derived from first principles. Furthermore, the distributional boundary potential added     
to the Hamiltonian is proportional to $\hbar ^2$ contrary to what is standard in the deformation context where typically the Hamiltonian is not modified by  quantum corrections (i.e. it is just the classical Hamiltonian).\\

4) On the other hand in \cite{Dias1} the confined stargenvalue equation is just the Weyl transform of a new confined eigenvalue equation valid in standard operator quantum mechanics. Several important results follow from this property. In \cite{Dias1} we were able to: i) supply the boundary conditions to be satisfied by the confined stargenfunctions (this are just the Weyl transform of the Dirichlet boundary conditions on the wave function), ii) extend Baker's converse construction to the confined case, iii) extend the formalism to non-diagonal stargenfunctions and iv) generate a dynamical (Moyal) equation for the confined particle using exactly the same modification of the Hamiltonian.\\

To summarize: The method of Kryukov and Walton can be derived from first principles (i.e. by quantizing a regular potential) while our own approach is mathematically more complete. In some sense the two approaches complement each other.

Hence an interesting result would be the proof of the equivalence of the two formulations. In fact the original aim of the authors of \cite{Walton} was "to derive the prescription of \cite{Dias1} from first principles". However, they found that their "results do not relate easily to the proposal of \cite{Dias1}".

In this paper we will prove the equivalence of the two formulations for the simplest case of a free particle confined to the negative half-line.  
We will also provide the necessary boundary and kinematical conditions for the approach of \cite{Walton}  (thus proving Baker's converse construction for the simplest case mentioned above). Unfortunately, we were unable to extend these results to the time dependent case. We discuss this issue in section 3.  

\section{Equivalence of the two formulations}

We shall now try to explain how equations (10) and (16) are related. The crux of the matter is the fact that (16) are equations for the Wigner function $f(x,p)$ whereas (10) is an equation for $\rho (x,p)$. The two quantities are related by: 
\begin{equation}
f(x,p) = \theta (-x) \rho(x,p).
\end{equation}
Following the method of \cite{Walton}, we get from eq.(16):
\begin{equation}
H \star f \star H = E^2 f \Longleftrightarrow p^2 \star f \star p^2 + p^2 \star f \star \delta_-'(x) + \delta_-'(x) \star f \star p^2 + \delta_-'(x) \star f \star \delta_-'(x) = E^2 f.
\end{equation}
From (16), we also have:
\begin{equation}    
\delta_-'(x) \star f = E f - p^2 \star f, \hspace{0.5 cm} f \star \delta_-'(x) = E f - f \star p^2.
\end{equation}
Substituting (20) in (19) we get:
\begin{equation}
2 E Re (p^2 \star f) - p^2 \star f \star p^2 + \delta_-'(x) \star f \star \delta_-'(x) = E^2 f.
\end{equation}
Now, as stated before, the solution of (16,17) takes the form (4) where the wavefunction obeys a Dirichlet boundary condition. In \cite{Dias1} we proved that $f (x,p)$ must then satisfy:
\begin{equation}
f (0^-,p) = \partial_x f (0^-,p) = \partial_x^2 f (0^-,p)=0 ,\quad \forall p \in \bkR,
\end{equation}
a result that follows from eq.(4) and which implies:
\begin{equation}
\rho (0,p) = \partial_x \rho (0,p) = \partial_x^2 \rho (0,p)=0, \hspace{0.5 cm} \forall p \in \bkR.
\end{equation}
From eq.(18) it follows that:
\begin{equation}
\partial_x f (x,p) =-\delta(x)\rho(x,p)+\theta (-x) \partial_x \rho (x,p)=
 \theta (-x) \partial_x \rho (x,p)
\end{equation}
where eq.(23) was taken into account. Likewise:
\begin{equation}
\left\{
\begin{array}{l}
\partial_x^2 f (x,p) = \theta (-x) \partial_x^2 \rho (x,p)\\
\partial_x^3 f (x,p) = \theta (-x) \partial_x^3 \rho (x,p)\\
\partial_x^4 f (x,p) = \theta (-x) \partial_x^4 \rho (x,p) - \delta (x) \partial_x^3 \rho (0,p)
\end{array}
\right.
\end{equation}
On the other hand, a simple calculation following \cite{Dias1} leads to:
\begin{equation}
\delta_-' (x) \star f (x,p) \star \delta_-' (x) = \frac{1}{2\pi}  \delta (x) \left| \psi ' (0) \right|^2.
\end{equation}
From (18,21,24,25,26), we get:
\begin{equation}
\begin{array}{c}
(2 E p^2 - p^4 ) \theta (-x) \rho - \frac{1}{2} ( E+ p^2) \theta (-x) \partial_x^2 \rho - \frac{1}{16} \left[\theta (-x) \partial_x^4 \rho (x,p) - \delta (x) \partial_x^3 \rho (0,p) \right] + \\
\\
+ \frac{1}{2 \pi} \delta (x)  \left| \psi ' (0) \right|^2 = E^2 \theta (-x) \rho (x,p).
\end{array}
\end{equation}
Finally, if we compute $\partial_x^3 \rho (0,p)$, using (4), we obtain:
\begin{equation}
\partial_x^3 \rho (0,p) = - \frac{8}{\pi} \left| \psi ' (0) \right|^2.
\end{equation}
Substituting (28) in (27) we obtain precisely (10). This means that our equation (16) together with the boundary condition (17) implies the main result of Kryukov and Walton (10).

We will now argue that the two approaches are in fact equivalent, provided we require that the solution of (10) be subjected to the boundary conditions (17,23) and to a pure state condition to be stated below \cite{Dias3}.

We start by looking for solutions of (10) of the form $e^{i r(p) x}$. From (10) we obtain the 4 solutions: $r_1 (p)  = r_2^* (p) = 2i (p + \sqrt E) $, $r_3 (p) = r_4^* (p) = 2i (p - \sqrt E) $. Therefore, the most general real solution of (10) is:
\begin{equation}
\rho (x,p) = A (p) e^{2i (p + \sqrt E)x} + A^* (p) e^{- 2i (p + \sqrt E)x} + B (p) e^{2i (p - \sqrt E)x} + B^* (p) e^{- 2i (p - \sqrt E)x}.
\end{equation}
Now, imposing the boundary conditions (23), which are necessary (albeit not sufficient) to ensure that $\psi (0) =0$ \cite{Dias1}, we get:
\begin{equation}
\rho (x,p) = N (p) \left\{ \frac{\sin \left[2x (p + \sqrt E) \right]}{p + \sqrt E} - \frac{\sin \left[2x (p - \sqrt E) \right]}{p - \sqrt E} \right\},
\end{equation}
where $N(p)$ is an arbitrary real function of $p$. Notice that if we choose
\begin{equation}
N(p) \propto \frac{1}{p},
\end{equation}
we obtain (up to a multiplicative real constant) the physical solution (eq.(30) of ref.\cite{Walton}). To derive (31) we still have to impose a pure state condition and condition (17). In ref.\cite{Dias3} we proved that if
\begin{equation}
\Sigma (y,p) \equiv \int dx \hspace{0.2 cm} e^{i xy} f (x,p),
\end{equation}
obeys the nonlinear partial differential equation
\begin{equation}
\frac{\partial^2}{\partial y^2} \ln \Sigma (y,p) = \frac{1}{4} \frac{\partial^2}{\partial p^2} \ln \Sigma (y,p),
\end{equation}
then $f(x,p)$ is a pure state Wigner function. Let us apply this condition to $f(x,p) = \theta (-x) \rho (x,p)$, where $\rho (x,p)$ is given by (30):
\begin{equation}
\ln \Sigma (y,p) = \ln \left[N (p) p \right] + \ln  \Sigma_1 (y,p),
\end{equation}
where:
\begin{equation}
\Sigma_1 (y,p) \equiv \int dx \hspace{0.2 cm} \theta (-x) \frac{e^{i xy}}{p}  \left\{ \frac{\sin \left[2x (p + \sqrt E) \right]}{p + \sqrt E} - \frac{\sin \left[2x (p - \sqrt E) \right]}{p - \sqrt E} \right\}.
\end{equation}
From the comments made after eq.(31), we already know that $\ln \Sigma_1 (y,p)$ satisfies (33). We are left with:
\begin{equation}
\frac{\partial^2}{\partial p^2} \ln \left[p N (p) \right] =0,
\end{equation}
the solution of which is
\begin{equation}
N(p) \propto \frac{e^{a p }}{p},
\end{equation}
where $a$ is an arbitrary real constant. We then get:
\begin{equation}
f(x,p) = \theta (-x) \rho (x,p) \propto \int_{- \infty}^{+ \infty} dy \hspace{0.2 cm} e^{- 2iyp} \varphi^* (x-y) \varphi (x+y),
\end{equation}
where 
\begin{equation}
\varphi (x+ia /2 ) = \theta (-x) \left(e^{i x \sqrt E} - e^{- i x \sqrt E} \right).
\end{equation}
Finally, if we require $f (x,p)$ to satisfy (17), we get $a=0$, which is the desired solution (and which also avoids the continuation of the Heaviside function into the complex plane). 

So, it seems that the problem can be formulated equivalently in the two following distinct ways. Either $f(x,p)$ is a solution of (16,17) or, alternatively, $f(x,p) = \theta (-x) \rho (x,p)$, where $\rho (x,p)$ is a solution of (10,17,23,32,33). 
The approach of Kryukov and Walton (as our own) remains valid if an additional non-singular potential is added to the Hamiltonian. We expect the equivalence of the two formulations to be extendable to this more general case. However this is still lacking a proper proof.

\section{Discussion}

From the previous analysis it seems that the two methods are equivalent provided suitable boundary and kinematical conditions are imposed. At least this seems to be the case for the time-independent stargenfunctions of non-degenerate Hamiltonians. 
By proving this equivalence for the time independent case we also provided a derivation of the approach of \cite{Dias1} from first principles. We also supplied the boundary and kinematical conditions selecting the confined stargenfunctions in the method of \cite{Walton} thus extending Baker's converse construction for this approach. Notice that these results still have to be extended to the case where an additional regular potential is added. 

On the other hand, the Kryukov-Walton method still lacks a suitable extension to the non-diagonal and to the time dependent cases. Unfortunately we think that a time-dependent equation is difficult to obtain with this approach. Indeed, the key to deriving eq.(10) is the fact that the stargenvalue equation has both a real and an imaginary part, which allow us to express $\partial_x \rho$ and $\partial_x^2 \rho$ in terms of the quantities $\rho (x, p \pm i \alpha)$ and obtain eq. (37) of ref.\cite{Walton}. However, the Moyal equation,
\begin{equation}
i \hbar \frac{\partial}{\partial t} \rho (x,p,t) =  H(x,p) \star \rho (x,p,t) - \rho (x,p,t) \star H (x,p),
\end{equation}
only has a real part. Hence, the previous method cannot be applied straightforwardly. 

Alternatively, we may follow the method presented here and start from the confined Moyal equation (eqs.(41,42) below). It is conceivable that one may eliminate the delta functions appearing in this equation. In \cite{Dias1} we proved that for the free particle it reads:
\begin{equation}
\partial_t \rho + 2 p \partial_x \rho = {\cal K} (x,p,t),
\end{equation}
where
\begin{equation}
{\cal K} (x,p,t) = \frac{1}{2i} \left[e^{-2ipx} \psi'^* (0,t) \psi (2x,t) - e^{2ipx} \psi^* (2x,t) \psi' (0,t) \right].
\end{equation}
Now suppose that the latter expression obeys the differential equation:
\begin{equation}
\partial_t {\cal K} = {\cal D} \left( \partial_x, \partial_p \right) {\cal K},
\end{equation}
where ${\cal D}$ is some differential operator. Then, from (41,43) we get:
\begin{equation}
\left(\partial_t - {\cal D} \left( \partial_x, \partial_p \right) \right)  \left( \partial_t \rho + 2 p \partial_x \rho \right) = 0.
\end{equation}
 
In any case, even if it were possible to obtain such an equation, the previous analysis reveals that this will most likely be a second-(or higher-)order equation with respect to time. This in turn spoils one of the most important features of deformation quantization, namely that the time-evolution should be a deformation of the classical Liouville equation.

\vspace{1 cm}

\begin{center}

{\large{{\bf Acknowledgments}}} 

\end{center}

\vspace{0.3 cm}
\noindent
This work was partially supported by the grant POCTI/MAT/45306/2002.


\begin{thebibliography}{10}


\bibitem{Walton} S.Kryukov, M.A.Walton, Ann. Phys. 317 (2005) 474.

\bibitem{Walton2} S.Kryukov, M.A.Walton, arxiv: quant-ph/0508005.

\bibitem{Dias1} N.C.Dias, J.N.Prata, J. Math. Phys. 43 (2002) 4602.

\bibitem{Dias2} N.C.Dias, J.N.Prata, Mod. Phys. Lett. A 20 (2005) 1371.

\bibitem{Piotr} P. Garbaczewski, W. Karwowski, Am. J. Phys. 72 (2004) 924.

\bibitem{Isham} C. Isham, {\it Topological and global aspects of quantum theory}, in: Les Houches, Session XL, eds. B.S. DeWitt and R. Stora (Elsevier, 1984).

\bibitem{Dias3} N.C.Dias, J.N.Prata, Ann. Phys. 313 (2004) 110.


\end{thebibliography}
\end{document}